\documentclass[aps,prl,twocolumn,groupedaddress,showpacs]{revtex4}
\usepackage{epsfig}
\usepackage[dvipsnames,usenames]{color}
\emergencystretch=\maxdimen
\begin{document}

\title{
Competing Supersolid and Haldane Insulator phases in 
the extended one-dimensional bosonic Hubbard model
}

\author{G.G. Batrouni$^{1,2,3}$, R.T.~Scalettar$^4$, V. G, Rousseau$^5$,
  B. Gr\'emaud$^{3,6,7}$}

\affiliation{$^1$INLN, Universit\'e de Nice-Sophia Antipolis, CNRS;
1361 route des Lucioles, 06560 Valbonne, France }

\affiliation{$^2$Institut Universitaire de France}

\affiliation{$^3$Centre for Quantum Technologies, National
University of Singapore; 2 Science Drive 3 Singapore 117542}

\affiliation{$^4$Physics Department, University of California, Davis,
California 95616, USA}

\affiliation{$^5$Department of Physics and Astronomy, Louisiana State
  University, B\^aton Rouge, Louisiana 70803, USA}

\affiliation{$^6$Laboratoire Kastler Brossel, UPMC-Paris 6, ENS, CNRS;
  4 Place Jussieu, F-75005 Paris, France}

\affiliation{$^7$Department of Physics, National University of
Singapore, 2 Science Drive 3, Singapore 117542, Singapore}

\begin{abstract}
  The Haldane Insulator is a gapped phase characterized by an exotic
  non-local order parameter.  The parameter regimes at which it might
  exist, and how it competes with alternate types of order, such as
  supersolid order, are still incompletely understood. Using the
  Stochastic Green Function (SGF) quantum Monte Carlo (QMC) and the
  Density Matrix Renormalization Group (DMRG), we study numerically
  the ground state phase diagram of the one-dimensional bosonic
  Hubbard model (BHM) with contact and near neighbor repulsive
  interactions. We show that, depending on the ratio of the near
  neighbor to contact interactions, this model exhibits charge density
  waves (CDW), superfluid (SF), supersolid (SS) and the recently
  identified Haldane insulating (HI) phases. We show that the HI
  exists only at the tip of the unit filling CDW lobe and that there
  is a stable SS phase over a very wide range of parameters.
\end{abstract}

\maketitle

Since its introduction in 1989 \cite{fisher89}, the bosonic Hubbard
model (BHM) has attracted continued interest due to its very rich
ground state phase diagram especially in the presence of longer range
interactions. Direct experimental relevance was established with the
realization of this model Hamiltonian, with tunable parameters
\cite{jaksch99}, in systems of ultra-cold bosonic atoms loaded in
optical lattices \cite{greiner02}. In its simplest form, a single
boson species with only contact repulsion, the system exhibits two
phases in the ground state \cite{fisher89}, a superfluid (SF) and an
incompressible Mott insulator (MI) depending on the particle filling
and the interaction strength. Extensive quantum Monte Carlo (QMC)
simulations have established that, when longer range interactions are
included, the supersolid (SS) phase can form for a wide range of
parameters and lattice geometries in one, two and three dimensions
\cite{batrouni95,batrouni00,goral02,wessel05,boninsegni05,sengupta05,otterlo05,batrouni06,yi07,suzuki07,dang08,pollet10,capogrosso10}. Typically,
the SS phase is produced by doping a phase exhibiting long range
charge density order (CDW). However, further QMC work has shown that,
remarkably, for a wide range of parameters, the extended BHM with near
neighbor interactions exhibits a SS phase even at {\it commensurate}
fillings in two and three dimensions \cite{kawashima12a,kawashima12b}.

In addition, it was shown that the one-dimensional extended BHM with
next near and/or near neighbor interactions admits another exotic
phase at a filling of one particle per site; the Haldane insulator
(HI) \cite{altman06,altman08}. The HI is a gapped insulating phase
characterized by a highly non-local order parameter like the Haldane
phase \cite{haldane83,dennijs89} in integer spin chain systems (see
below). This gives rise to several questions. Does the HI exist for
other integer fillings of the system or is it a special property of
the unit filling case? The SS phase found in one dimension
\cite{batrouni06} was obtained by doping a CDW phase: Does this phase
also exist for commensurate fillings in one dimension for parameter
choices similar to those in two \cite{kawashima12a} and three
dimensions \cite{kawashima12b}? If the SS phase exists for
commensurate fillings, where is it situated in the phase diagram
relative to the CDW, MI and HI phases? The phase diagram at unit
filling for the BHM with contact ($U$) and near neighbor ($V$)
interactions was determined via QMC \cite{batrouni91,batrouni94} and
found to have SF, MI and CDW phases but no SS. Subsequently, the
$(\mu,t)$ phase diagram of the extended BHM, for a fixed $V/U$ ratio,
was obtained using Density Matrix Renormalization Group
(DMRG)~\cite{kuhner00}, but showed only evidence for MI, SF and
CDW. More recent work, also based on the DMRG, has found
\cite{rossini12} no SS phase in the $(U,V)$ plane at unit filling but
the question of other fillings was not addressed. Reference
\cite{rossini12} also found a HI phase sandwiched between the MI and
CDW phases which was not present in \cite{batrouni91,batrouni94}. As
we shall see below, the HI phase was not found in the earlier work
because the superfluid density in this phase vanishes very slowly with
the system size and the largest sizes that could be accessed at the
time were $64$ sites.

Theoretical studies of this system using bosonization have led to
mixed results. The HI was obtained and characterized with bosonization
\cite{altman08} but consensus is absent on whether the SS phase exists
in this model. Even though older studies did not specifically mention
it~\cite{Giamarchi_book} or even argued that it did not
exist~\cite{kuhner00}, more recent studies seem to demonstrate the
presence of the SS phase ~\cite{Sengupta07}, even without nearest
neighbor interaction~\cite{Lazarides11}, for both commensurate and
incommensurate fillings. However, the precise nature of order and the
decays of the relevant correlation functions are still far from
settled.  For instance, some studies predict that the single particle
Green function decays exponentially in the SS phase while the
density-density correlation function decays as a power \cite{Lee07};
others predict that both of these correlation functions decay as
powers~\cite{Sengupta07}. Finally, the universality class of the
transition to the SS phase remains largely unexplored.

In this Letter we answer some of these questions using the Stochastic
Green Function (SGF) QMC algorithm \cite{sgf} and the ALPS \cite{alps}
DMRG code to obtain the phase diagram of the extended BHM in one
dimension,
\begin{eqnarray}
\nonumber
 H &=& -t\sum_{i} (a^{\dagger}_ia^{\phantom\dagger}_{i+1} +
a^{\dagger}_{i+1}a^{\phantom\dagger}_{i}) + \frac{U}{2} \sum_i
n_i\left(n_i-1\right)\\
&& + V\sum_in_in_{i+1}.
\label{ham}
\end{eqnarray}
The sum over $i$ extends over the $L$ sites of the lattice, periodic
boundary conditions were used in the QMC and open conditions with the
DMRG. The hopping parameter, $t$, is put equal to unity and sets the
energy scale, $a_i$ ($a_i^\dagger$) destroys (creates) a boson on site
$i$, $n_i= a_i^\dagger a_i$ is the number operator on site $i$, $U$
and $V$ are the onsite and near neighbor interaction parameters. All
results presented here were obtained at the fixed ratio $V/U=3/4$
which favors CDW phases over MI at commensurate fillings when $U$ is
large.

\begin{figure}[t]
\centerline{\epsfig{figure=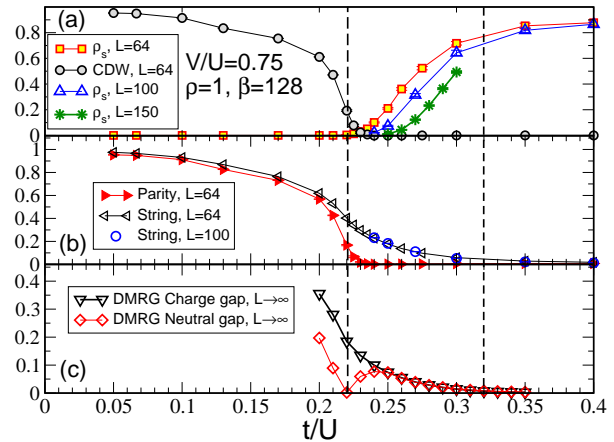,width=9.5cm,clip}}
\caption{(color online) Shows several quantities for $\rho=1$ as
  functions of $t/U$ with the fixed ratio $V/U=0.75$. (a) The CDW order
  parameter for $L=64$ and $\rho_s$ for $L=64,\, 100,\, 150$; (b) the
  parity and string order parameters; (c) the neutral and charge
  gaps. (a) and (b) were obtained with QMC and (c) with DMRG. The
  region between the vertical dashed lines is the HI. In the HI,
  $\rho_s\to 0$ very slowly as $L$ increases while the string
  parameter is essentially constant. Note the difference between the
  neutral and charge gaps. The gaps are given in units of the hopping
  $t$.
  \label{fig1} }
\end{figure}

Several quantities are needed to characterize the phase diagram. The
superfluid density is obtained with QMC using \cite{pollock}
\begin{equation}
\rho_s = \frac{\langle W^2\rangle}{2td\beta L^{d-2}},
\label{rhos}
\end{equation}
where $W$ is the winding number of the boson world lines, $d$ is the
dimensionality and $\beta$ the inverse temperature. The CDW order
parameter is the structure factor, $S(k)$, at $k=\pi$ where
\begin{equation}
S(k) = \frac{1}{L}\sum_{r=0}^{L-1} {\rm e}^{ikr}\langle n_0n_r
\rangle,
\label{sk}
\end{equation}
and the momentum distribution, $n_k$, is given by
\begin{equation}
n_k=\sum_{r=0}^{L-1} {\rm e}^{ikr}\langle a^{\dagger}_0
a^{\phantom\dagger}_r \rangle.
\label{nk}
\end{equation}
The charge gap is given by $\Delta_c(n)=\mu(n)-\mu(n-1)$; the chemical
potential is $\mu(n)=E_0(n+1)-E_0(n)$ where $E_0(n)$ is the ground
state energy of the system with $n$ particles and is obtained both
with QMC and DMRG. The neutral gap, $\Delta_n$, is obtained using DMRG
by targeting the lowest excitation with the same number of bosons. In
both CDW and HI phases, the chemical potentials at both ends are set
to (opposite) large enough values, in DMRG, such that the ground state
degeneracy and the low energy edge excitations are
lifted~\cite{kuhner00,altman06}.  With the SGF we did simulations in
both the canonical and grand canonical ensembles.

For large values of $U$ and $V$ at $\rho=1$, a site typically has
$n_r=0,1,2$ particles with higher occupations being very rare. The
system then becomes analogous to an $S=1$ spin chain \cite{altman06}
with $S_z(i)\equiv \delta n_i=n_i-\rho$ ($\rho=1$ here) taking values
of $0,\pm 1$. Consequently, string and parity operators can be defined
\cite{altman06,altman08} to characterize the Haldane Insuating phase,
\begin{eqnarray}
{\cal O}_s(|i-j|) &=& \langle \delta n_i {\rm e}^{i\theta \sum_{k=i}^j
  \delta n_k} \delta n_j \rangle,\\
\label{string}
{\cal O}_p(|i-j|) &=& \langle  {\rm e}^{i\theta \sum_{k=i}^j \delta
  n_k} \rangle,
\label{parity}
\end{eqnarray}
where $\theta=\pi$ for $S=1$. The corresponding value of the order
parameter is obtained in the limit $|i-j| \to \infty$; in practice we
take the order parameters to be ${\cal O}_{s/p}(L_{max})$ where, in
QMC with PBC, $L_{max}=L/2$ and in DMRG, with OBC, $L_{max}$ is the
longest distance possible before edge effects start being felt. For
higher integer filling, $\rho=2,3\dots$, $\theta\neq \pi$ and has to
be determined as discussed in \cite{qin03}. It was shown
\cite{altman06,altman08} that the phases are characterized as follows
\cite{foot1}: In MI, $\rho_s=0$, $S(\pi)=0$, $\Delta_c=\Delta_n \neq
0$, ${\cal O}_p(L_{max}) \neq 0$, ${\cal O}_s(L_{max})= 0$; in CDW,
$\rho_s=0$, $S(\pi)\neq 0$, $\Delta_c$ and $\Delta_n \neq 0$, ${\cal
  O}_p(L_{max})$ and ${\cal O}_s(L_{max})\neq 0$; in SF $\rho_s\neq
0$, $S(\pi)=0$, $\Delta_c=\Delta_n= 0$, ${\cal O}_p(L_{max})={\cal
  O}_s(L_{max})= 0$; in HI, $\rho_s=0$, $S(\pi)=0$, $\Delta_c$ and
$\Delta_n \neq 0$, ${\cal O}_p(L_{max})= 0$, ${\cal O}_s(L_{max})\neq
0$. In the SS phase, CDW and SF order coexist and one expects
$\rho_s\neq 0$, $S(\pi)\neq 0$, $\Delta_c=\Delta_n = 0$ and, because
$S(\pi)\neq 0$, ${\cal O}_p(L_{max}) \neq 0$, ${\cal O}_s(L_{max})\neq
0$. The gaps, $\Delta_c$ and $\Delta_n$ behave in a subtle way in the
CDW and HI phases (see below).

\begin{figure}[t]
\centerline{\epsfig{figure=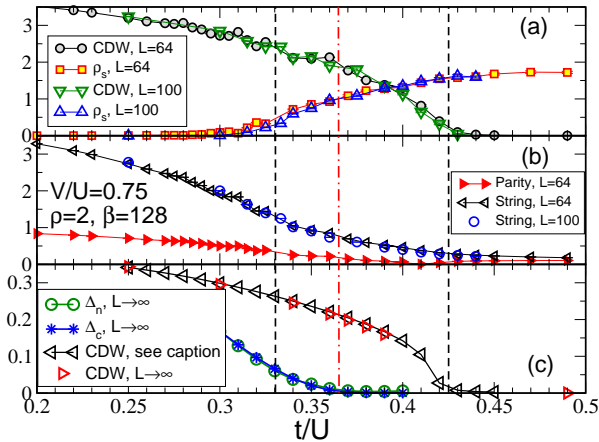,width=9.5cm,clip}}
\caption{(color online) Same as Fig.~\ref{fig1} but at $\rho=2$. (a)
  QMC simulations show that in the interval between the two vertical
  (black) dashed lines there is similtaneous SF and CDW order and,
  therefore, a supersolid phase (SS). The vertical (red) dot-dash
  line is where the $L\to \infty$ extrapolated neutral ($\Delta_n$)
  and charge ($\Delta_c$) gaps vanish in DMRG (c). The CDW-SS
  transition is between $t/U=0.33$ (QMC) and $t/U=0.355$
  (DMRG). The difference between the two values could be due to the
  difference in the boundary conditions, open for DMRG and periodic
  for QMC. (c) also shows the $L\to \infty$ extrapolated CDW order
  parameter, right (red) triangles, and the Fourier transform of
  $\langle n_i\rangle\langle n_j\rangle $, left (black) triangles,
  obtained with DMRG to probe the disappearance of CDW order. Both DMRG
  and QMC give the SS-SF transition at $t/U\approx 0.425$.  Note that,
  unlike Fig.~\ref{fig1}, the charge and neutral gaps (c) are
  essentially always the same. 
  \label{fig2} }
\end{figure}

We start with the system at $\rho=1$ and study the phases as functions
of $t/U$. Figure \ref{fig1} shows the dependence on $t/U$ of $\rho_s$,
$S(k=\pi)$, ${\cal O}_s$, ${\cal O}_p$, $\Delta_c$ and
$\Delta_n$. Figure \ref{fig1}(c) shows that in the CDW phase $\Delta_c
> \Delta_n$ and that $\Delta_n=0, \,\Delta_c\neq 0$ at the CDW-HI
transition. The HI-SF transition is signaled by $\Delta_c=\Delta_n \to
0$ \cite{altman06,altman08}. Finite size scaling of the DMRG results
show that $\Delta_n\to 0$ at $t/U\approx 0.32\pm 0.01$. Therefore,
according to the criteria discussed above, the system is in the CDW
phase for $t/U\leq 0.22$ and in the SF phase for $t/U \geq 0.32$. For
$0.22 \leq t/U \leq 0.32$ (between the two vertical dashed lines), the
system is in the HI phase with $\rho_s \to 0$ as the size
increases. Note how slowly $\rho_s\to 0$ with increasing $L$ and how
insensitive ${\cal O}_s$ is to the finite size. This makes ${\cal
  O}_s$ a more reliable indicator at moderate system sizes.  Our
CDW-HI transition at $t/U=0.22$ agrees very well with Fig.~1 in
\cite{rossini12}. However, the value we obtain for the HI-SF
transition, $t/U\approx 0.32$ does not agree with the schematic dashed
line in that figure.

As mentioned above, the behavior at $\rho=1$ may be understood by
making the analogy with $S=1$ spin chains. The question arises then as
to whether such an analogy between this extended BHM at
$\rho=2,3\dots$ and $S=2, 3\dots$ spin chains is valid and also leads
to HI phases. This question is addressed in Fig.~\ref{fig2} for
$\rho=2$ which shows qualitatively different behavior compared to
Fig.~\ref{fig1}. While for low $t/U$ both cases exhibit CDW phases,
the behavior of $\Delta_c$ and $\Delta_n$, calculated with DMRG, is
strikingly different as seen in Fig.~\ref{fig2}(c): For $\rho=2$,
$\Delta_c=\Delta_n$ and finite size scaling shows that they vanish
together at $t/U\approx 0.36$, which indicates that there
  is no HI in this case. This is consistent with the absence of the
  Haldane phase in $S=2$ spin chains \cite{evenS}. Nonetheless, for
this filling, the system does exhibit another salient feature: Indeed,
figure \ref{fig2}(a) and (c) show from both QMC and DMRG that when the
gaps vanish, $S(\pi)$ remains non-zero while $\rho_s$ also takes a
non-zero value. $S(\pi)$ and $\rho_s$ both remain non-zero for
$0.33\leq t/U \leq 0.425$ indicating the presence of a supersolid
phase.  The CDW-SS transition is estimated to be at $t/U\approx 0.33$
from QMC and $t/U\approx 0.36$ from DMRG while both DMRG and QMC give
$t/U\approx 0.425$ for the SS-SF transition.  In Fig.~\ref{fig4} we
show $n_k/L$ and $S(k)$ in the SS phase at $t/U=0.35$ and $L=64, 100,
128$. We see that while $n_k/L\to 0$ (see center peak, $k=0$) as
expected (since there is no condensate in one dimension), the peaks in
$S(k)$ do not depend on $L$, indicating long range CDW order. This
behavior is also confirmed by a finite-size scaling analysis of the
DMRG results for sizes $L=64, 96, 128, 160$.  For the three phases
(CDW, SS and SF), the CDW order parameter $S(\pi)$ is found to scale
as $S_0+S_1/L+S_2/L^2$, whereas $n_0/L$ is always found to decay as a
power law $n_1/L^{\alpha}$.  More precisely, one obtains the following
results: for $t/U=0.25$ (CDW): $S_0=3.4$ and $\alpha\approx 0.95$; for
$t/U=0.39$ (SS): $S_0=1.6$ and $\alpha\approx 0.22$; for $t/U=0.49$
(SF): $S_0\approx0$ and $\alpha\approx 0.14$.  Note that, in both $SS$
and $SF$ phases, the parameter $\alpha$ is less than $0.25$, in
agreement with a Luttinger liquid description of the
system~\cite{Giamarchi_book,altman06,altman08,Sengupta07}.  This
scaling law and the insensitivity of $\rho_s$ to $L$,
Fig.~\ref{fig2}(a), confirm that this is indeed the SS phase. This
surprising appearance of the SS phase at commensurate filling has also
been observed in two and three dimensions
\cite{kawashima12a,kawashima12b}. Furthermore, we find that the Green
function, $G(r) = \langle a^\dagger_r a^{\phantom\dagger}_0\rangle$,
decays as a power in the SS phase with exponent $\approx 0.5$ at $t/U
= 0.34$.

For the present value of $V/U$, this behavior at $\rho=2$ is repeated
at $\rho=3$ (and presumably at higher integer fillings): As $t/U$ is
increased, the system goes from CDW to SS to SF without exhibiting any
HI phases.  It appears, therefore, that, at least for $V/U=3/4$, the
analogy between integer spin chains and this extended BHM at integer
fillings applies only at $\rho=1$.

\begin{figure}[t]
\centerline{\epsfig{figure=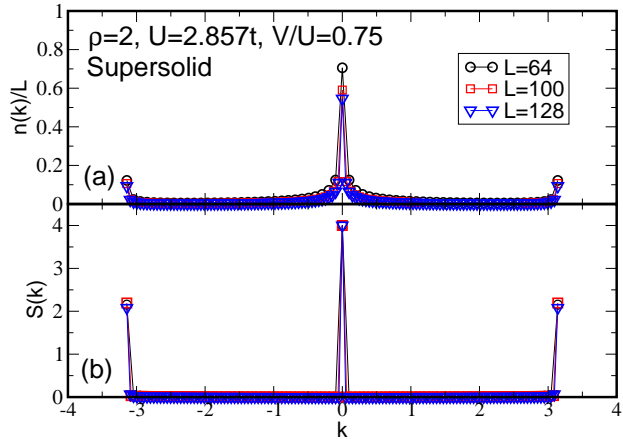,width=9.5cm,clip}}
\caption{(color online) The dependence of the momentum distribution,
  $n_k/L$ and the structure factor,$S(k)$ on the system
  size. $n_k/L \to 0$ with increasing $L$ while $S(k)$ remains
  constant indiating long range CDW order.
\label{fig3} }
\end{figure}

\begin{figure}[t]
\centerline{\epsfig{figure=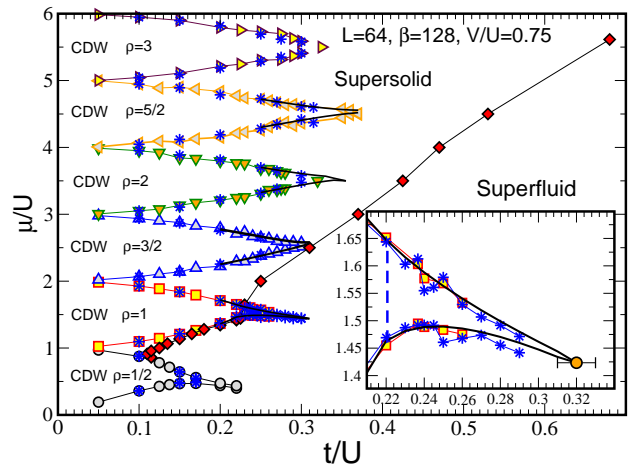,width=9.5cm,clip}}
\caption{(color online) Main panel: The phase diagram of the BHM
  obtained with QMC and DMRG simulations. All results are for a system
  with $V/U=0.75$. All symbols are from QMC and are for $L=64$ sites,
  $\beta=128$ (stars are for $L=128$). Black lines near the tips of
  the CDW lobes are DMRG results for $L=192$. Inset: A zoom of the tip
  of the $\rho=1$ lobe. To the right of the vertical dashed line
  ($t/U=0.22$) is the HI phase, to the left is the CDW.
    \label{fig4} }
\end{figure}

The phase diagram in the ($t/U,\mu/U$) plane for $V/U=0.75$ is mapped
by calculating the charge gaps at commensurate fillings, multiples of
$L/2$, and by making plots like Figs.~\ref{fig1}, \ref{fig2} which
allow us to identify the various phases.  We have also verified that,
for incommensurate fillings, DMRG agrees with QMC on the boundary
between the SS-SF.  The result is shown in Fig.~\ref{fig4}. Figure
\ref{fig4} was obtained using QMC (all symbols) and DMRG (black lines
near lobe tips). The QMC results are for $L=128$ (stars) and $L=64$
(all other symbols), $\beta=128$. The solid black lines near the lobe
tips are obtained from DMRG with $L=192$. The end points of the lobes
are obtained by studying the finite size dependence of $\Delta_n$
using DMRG. The inset is a zoom of the tip of the $\rho=1$ lobe.

Several comments are in order. The $\rho=1/2$ lobe is surrounded
almost entirely by SF except for a small region of SS squeezed between
it and the $\rho=1$ lobe. The fact that in the extended BHM a SS does
not exist when the $\rho=1/2$ CDW phase is doped with holes, but does
when it is doped with particles, was already addressed in
\cite{batrouni06}. The $\rho=1$ lobe sticks out of the SS phase and
the part sticking out is, in fact, the HI phase. No other CDW lobe
behaves this way. The $\rho=3/2$ lobe terminates right at the boundary
with the SF phase: To within the resolution of our simulations, the
transition from the $\rho=3/2$ CDW lobe goes directly into the SF
phase without passing through the SS phase.  This peculiar behavior
for $\rho=3/2$ was also observed with additional DMRG results for
different values of $V/U$ ranging from $0.65$ to $1$: The SS layer
between the CDW and SF phases, if present, is too thin to observe for
the considered system sizes. An accurate determination of the $(U,V)$
phase diagram for this filling will require a more thorough finite
size scaling analysis.  All other CDW lobes, $\rho\geq 2$, are
surrounded entirely by the SS phase. It is interesting to compare this
figure with Fig.~3 of \cite{kawashima12a} and with the mean-field
predictions~\cite{Iskin}.

In this letter we examined the phase diagram of the extended BHM at a
fixed ratio of the interaction terms, $V/U=3/4$. Contrary to
expectation, we found that this model at integer fillings does not
always behave analogously to integer spin chains. In particular, only
for $\rho=1$ and at small $t/U$ does this happen and the system
exhibits CDW, HI and SF phases. In the CDW phase at this filling,
$\Delta_c>\Delta_n$. At all other integer fillings, we found the HI
phase to be absent and in its place a supersolid phase which indicates
that the system at these fillings may not behave like an integer spin
chain. Furthermore, for all CDW phases, except the one at $\rho=1$, we
found that $\Delta_n=\Delta_c$ and that, unlike the $\rho=1$ case,
both gaps vanish together as the CDW phase gives way to SS or SF. It
is possible that, for a different $V/U$ ratio, the SS-SF boundary will
shift and cut the $\rho=3$ lobe (as it does in Fig. \ref{fig4} with
the $\rho=1$ lobe) resulting in a HI phase. If this happens, it could
mean there are two types of $\rho=3$ CDW phases, one in which the
neutral and charge gaps are always the same (what we find here) and
another CDW phase in which $\Delta_c>\Delta_n$ as is the case for the
$\rho=1$ CDW. We have also shown that the single particle Green
function decays as a power law in the SS phase.  Finally, from a
theoretical point of view, it would be interesting to characterize the
universality classes of the different quantum phase transitions
(CDW-SS-SF) occuring in the system.  In addition, in order to have
clear experimental signatures of the phases, in particular with cold
atom gases which allow time-resolved measurements one would need to
study the excitations of the system.

\acknowledgments We thank T. Giamarchi for very helpful
discussions. This work was supported by: the CNRS-UC Davis EPOCAL
joint research grant; by the France-Singapore Merlion program (PHC
Egide and FermiCold 2.01.09); by the LIA FSQL; by an ARO Award
W911NF0710576 with funds from the DARPA OLE Program and by NSF grant
OISE-0952300. The Centre for Quantum Technologies is a Research Centre
of Excellence funded by the Ministry of Education and National
Research Foundation of Singapore.



\begin{thebibliography}{50}
\bibitem{foot1} Note that our definitions of ${\cal O}_p$ and ${\cal
    O}_s$ differ slightly from \cite{altman06,altman08} in that the
  sum in the exponents is done from $i$ to $j$. This makes $|{\cal
    O}_p|=|{\cal O}_s|$ in the CDW phase for $\rho=1$.
\end{thebibliography}
\end{document}